\documentclass{aa}
\begin{document}

\title{Radio and X-ray images of the SNR G18.8+0.3 interacting with molecular clouds}

\author{W.W. Tian  \inst{1,2} 
\and
      D.A. Leahy \inst{2} 
\and
      Q.D. Wang \inst{3}}
\authorrunning{Tian, Leahy \& Wang}
\titlerunning{Radio and X-ray images of SNR G18.8+0.3}
\offprints{Wenwu Tian}
\institute{National Astronomical Observatories, CAS, Beijing 100012, China; email: tww@iras.ucalgary.ca\\
\and
Department of Physics \& Astronomy, University of Calgary, Calgary, Alberta T2N 1N4, Canada\\
\and
Department of Astronomy, University of Massachusetts, 710 North Pleasant Street, Amherst, MA 01003, USA}

\date{received xx, accepted xx, 2007} 
 
\abstract{
New HI images from the VLA Galactic Plane Survey (VGPS) show clear absorption features associated with the supernova remnant (SNR) G18.8+0.3.
High-resolution $^{13}$CO images reveal that molecular clouds overlap the radio filaments of G18.8+0.3. 
The $^{13}$CO emission spectrum over the full velocity range in the direction of G18.8+0.3 shows two molecular components with high brightness-temperature
and three molecular components with low brightness-temperature,
all with respective HI absorption. This implies that these clouds are in front of G18.8+0.3. In the HI images, the highest velocity absorption feature seen against the continuum image of G18.8+0.3 is at 129 km s$^{-1}$, which corresponds to the tangent point in this direction. This yields a lower distance limit of 6.9 kpc for G18.8+0.3. Absence of absorption at negative velocities gives an upper distance limit of 15 kpc. The broadened profile at 20$\pm$5 km s$^{-1}$ in the $^{13}$CO emission spectra is a strong indicator of a possible SNR/CO cloud interaction. Thus, G18.8+0.3 is likely to be at the distance of about 12 kpc. The upper mass limit and mean density of the giant CO cloud at 20$\pm$5 km s$^{-1}$ are $\sim$6 $\times$$10^{5}$$M_{\odot}$ and $\sim$ 2$\times$ 10$^{2}$ cm$^{-3}$. We find an atomic hydrogen column density in front of G18.8+0.3 of $N_{HI}$ $\sim$ 2$\times$10$^{22}$ cm$^{-2}$. 
The ROSAT PSPC observations show a diffuse X-ray enhancement apparently associated with part of the radio shell of G18.8+0.3.  
Assuming an association, the unabsorbed flux is 2.2$\times$10$^{-12}$ erg cm$^{-2}$s$^{-1}$, suggesting an intrinsic luminosity of 3.6$\times$10$^{34}$ erg s$^{-1}$ for G18.8+0.3.

}
\maketitle

\section{Introduction}
The interaction between a supernova remnant (SNR) and the surrounding interstellar molecular clouds can produce several observable physical phenomena, e.g. OH (1720 MHz) shock-excited masers (Frail et al. 1994; Frail \& Mitchell 1998; Wardle \& Yusef-Zadeh 2002), very high energy $\gamma$-rays (Torres et al. 2003; Tian et al. 2007) and puzzling X-ray morphological features (e.g. the lobe: the X-ray-bright interior feature in SNR CTB 109, Sasaki et al. 2006). The SNR/molecular cloud interaction may also potentially lead to new generations of star formation. SNR G18.8+0.3 (Kes 67) has been widely studied in radio bands including continuum, HI, CO and OH lines (Caswell et al. 1975; Kassim 1992; Dubner et al. 1996). Using morphological association between a giant molecular cloud (GMC) and G18.8+0.3, the SNR has been suggested to be interacting with the GMC (Dubner et al. 1999; 2004). 
Observations searching for OH lines (1720 MHz) and high energy $\gamma$-rays (Aharonian et al. 2006), have given no detection in the direction of G18.8+0.3. These factors have motivated us to study G18.8+0.3 further. In this paper, we use new radio 1420 MHz continuum and HI-line observations, high-resolution $^{13}$CO mapping, and unpublished X-ray data in order to provide new estimates for associated physical parameters such as the distance and intrinsic X-ray luminosity of G18.8+0.3, and to study the relation between G18.8+0.3 and the surrounding interstellar molecular and atomic clouds.   

\section{Observations}
\subsection{Radio Data}
The radio continuum at 1420 MHz and HI emission data sets come from the Very Large Array (VLA) Galactic Plane Survey (VGPS), described in detail by Stil et al. (2006). The continuum images of SNR G18.8+0.3 shown in this paper have a spatial resolution of 1$^{\prime}$ (FWHM) at 1420 MHz and an rms sensitivity of 0.3 K (T$_{b}/S_{\it{v}}$ =168 K Jy$^{-1}$). The synthesized beam for the HI line images is also 1$^{\prime}$, the radial velocity resolution is 1.56 km s$^{-1}$, and the rms noise is 2 K per channel (i.e. 0.824 km s$^{-1}$). The short-spacing information for the HI spectral line images comes from additional observations performed with the 100 m Green Bank Telescope of the NRAO.
The $^{13}$CO-line ($J$=1--0) data sets are from the Galactic ring survey (Jackson et al. 2006) carried out with the Five College Radio Astronomy Observatory 14 m telescope. The CO-line data sets in the paper have a velocity coverage of --5 to 135 km s$^{-1}$, an angular sampling of 22$^{\prime\prime}$, a radial velocity resolution of 0.21 km s$^{-1}$, and a rms noise of $\sim$ 0.13 K. 

\subsection {X-ray Data}
G18.8+0.3 was observed with the Position Sensitive Proportional Counter (PSPC) on the ROSAT X-ray Observatory on 1991 Mar. 30 and 1992 Apr. 2 (WG500048P-0 and WG500048P-1) for a total source integration time of 2805 and 1241 seconds respectively, and then between 1992 Oct. 10 and Oct. 13 (WG500211P) for 11057 seconds. In this work, we have combined all three observations. We present the exposure-corrected intensity image
in the 0.5-2.4 keV band; the intensity at lower energies is completely local in the solar neighborhood, because of interstellar absorption.
We have used the CIAO csmooth routine to obtain the combined images of G18.8+0.3, which is adaptively smoothed to have a S/N $\sim$ 3.  
A total of 13 point sources are detected in the 0.5$^{0}$ radius centered at G18.8+0.4 (see http://heasarc.gsfc.nasa.gov/W3Browse/ rosat/rospspctotal.html).  

\section{Results}
\subsection{Continuum Emission}

The VGPS continuum image of G18.8+0.3 at 1420 MHz is shown in Fig. 1. The VGPS map has a higher resolution and sensitivity (by factor of 3.5), and shows more details than the previous best image at 1465 MHz (Dubner et al. 1996). Additional faint features are detected by the present observations. The new image gives a corrected angular size of G18.8+0.3: 18$^{\prime}$$\times$13$^{\prime}$ in $l$ and $b$ directions. The radio continuum image of G18.8+0.3 looks rather as if there is a ``blow-out" towards the northwest (e.g. less well-defined shell structure, while the diffuse emission extends considerably). This is likely the region where the SNR shell is encountering lower densities (i.e., the molecular cloud is seen in the south edge in Fig. 5) and hence where a ``blow out" may be more likely.

\subsection{Producing HI Emission and Absorption Spectra}
We have searched the VGPS data within the radial velocity range from --113 to 165 km s$^{-1}$ for features in the HI emission which might be related to the morphology of G18.8+0.3. There are HI features closely coincident with the SNR in four velocity ranges: 
 0 to 8, 14 to 27, 48 to 56, and 100 to 116 km s$^{-1}$. Fig. 2 shows maps of HI emission for 4 channels: 5.13, 18.32, 54.60 and 113.96 km s$^{-1}$ respectively. These maps have superimposed contours (28, 48, 68, 80, 90, 100 K) of 1420 MHz continuum emission chosen to show the SNR. The HI features are strongly correlated with the continuum intensity, indicating that they are caused by absorption in HI between the SNR and the Earth.
\begin{table}
\begin{center}
\caption{Parameters of Six HI Spectra. }
\setlength{\tabcolsep}{1mm}
\begin{tabular}{ccccccc}
\hline
 Region:  &  1 & 2 & 3 & 4 & 5 & 6 \\
\hline
 No. source spectra & 72   & 63   &  42  & 66   & 90 & 20\\
 No. background spectra & 132  & 135  & 112  & 134  & 250& 90\\
 $T^{c}_{s}$(K)  & 57.1 & 63.8 & 76.0 & 68.8 & 56.2   &36.7\\
 $T^{c}_{bg}$(K)  & 35.6 & 33.9 & 38.3 & 35.5 & 34.3  & 28.0\\
\hline
\hline
\end{tabular}
\end{center}
\end{table}

We construct the HI emission and absorption spectra of G18.8+0.3 using a new method. 
Because we have an extended continuum source rather than a point source, we don't use the standard formula for HI absorption spectra. The continuum emission extends into our background region since the background region is chosen to be nearby the continuum peak in order to minimize the potential difference in the HI distribution along the two lines of sight (source and background). For the source line-of-sight: $T_{on}(\it{v})$ = T$_{B}(\it{v})$(1 - e$^{-\tau_{t}(\it{v})}$) + T$_{s}^{c}$(e$^{-\tau_{c}(\it{v})}$ - 1).
The HI images have continuum emission subtracted,  resulting in the ``-1" in the second term.  For the background line-of-sight: $T_{off}(\it{v})$ = T$_{B}(\it{v})$(1 - e$^{-\tau_{t}(\it{v})}$) + T$_{bg}^{c}$($e^{-\tau_{c}(\it{v})}$ - 1).
  This gives: $\Delta T$ = $T_{off}(\it{v})$ - $T_{on}(\it{v})$ = ($T^{c}_{s}$-$T^{c}_{bg}$)(1-$e^{-\tau_{c}(\it{v})}$). In above formulae, $T_{on}(\it{v})$ and  $T_{off}(\it{v})$ are the average brightness temperatures of many spectra from a selected area on a strong continuum emission region of the SNR and an adjacent background region.  $T^{c}_{s}$ and $T^{c}_{bg}$ are the average continuum brightness temperatures for the same regions respectively. $\tau_{t}(\it{v})$ is the total optical depth along line-of-sight and $\tau_{c}(\it{v})$ is the optical depth from the continuum source to the observer.
 When there is no continuum emission in the background region, this reduces to the standard formula (for continuum subtracted maps). 

Fig. 3 shows the source and background HI emission spectra and the absorption spectra for six selected regions on the face of G18.8+0.3. 
Each on-source region is outlined by a
solid-line box, within which the $T_{on}(v)$ is averaged. The respective
off-source region, within which the $T_{off}(v)$ is obtained, is defined by
the dashed-line box, excluding the solid boxed area. 
$T^{c}_{s}$ and $T^{c}_{bg}$ are measured from 1420 MHz continuum image by same method. Table 1 summarizes their parameters. Fig. 3 clearly reveals absorption features at 1, 5, 11, 18, 22, 27, 42, 51, 98, 113, 126 km s$^{-1}$. Region 3 contains the strongest continuum intensity on the face of G18.8+0.3 and thus shows the strongest absorption in the on-source spectrum (upper panel). There is little absorption at 60, 82, 93, 103 km s$^{-1}$, suggesting that little HI exists at these velocities: i.e. they probably correspond to interarm regions. Since in all 6 plots we see absorption features all the way up to near the tangential velocity at 130 km s$^{-1}$, the SNR must lie close to, or beyond, the tangent point in this direction.

All six regions show similar strong absorption features. This confirms that these are real and not due to spatial variation in the HI emission. The absorption spectra of region 6 (i.e. the protrusion, Dubner et al. 1999) is noisier due to lower continuum brightness temperature. The protrusion shows the same strong absorption features as the other regions, so it is likely part of G18.8+0.3. However, the protrusion almost looks like a separate discrete source in Fig. 1. It shows similar absorption to the SNR, another possibility is that it is a chance extragalactic source seen right through the galactic disc. 

\begin{figure}
\vspace{45mm}
\begin{picture}(60,60)
\put(-30,-120){\includegraphics{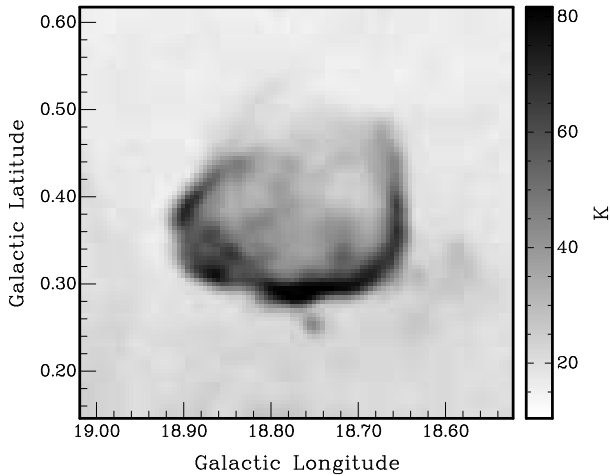}}
\end{picture}
\caption{The figure shows the VGPS continuum image of G18.8+0.3 at 1420 MHz.}
\end{figure}

\begin{figure*}
\vspace{120mm}
\begin{picture}(80,80)
\put(-10,110){\includegraphics{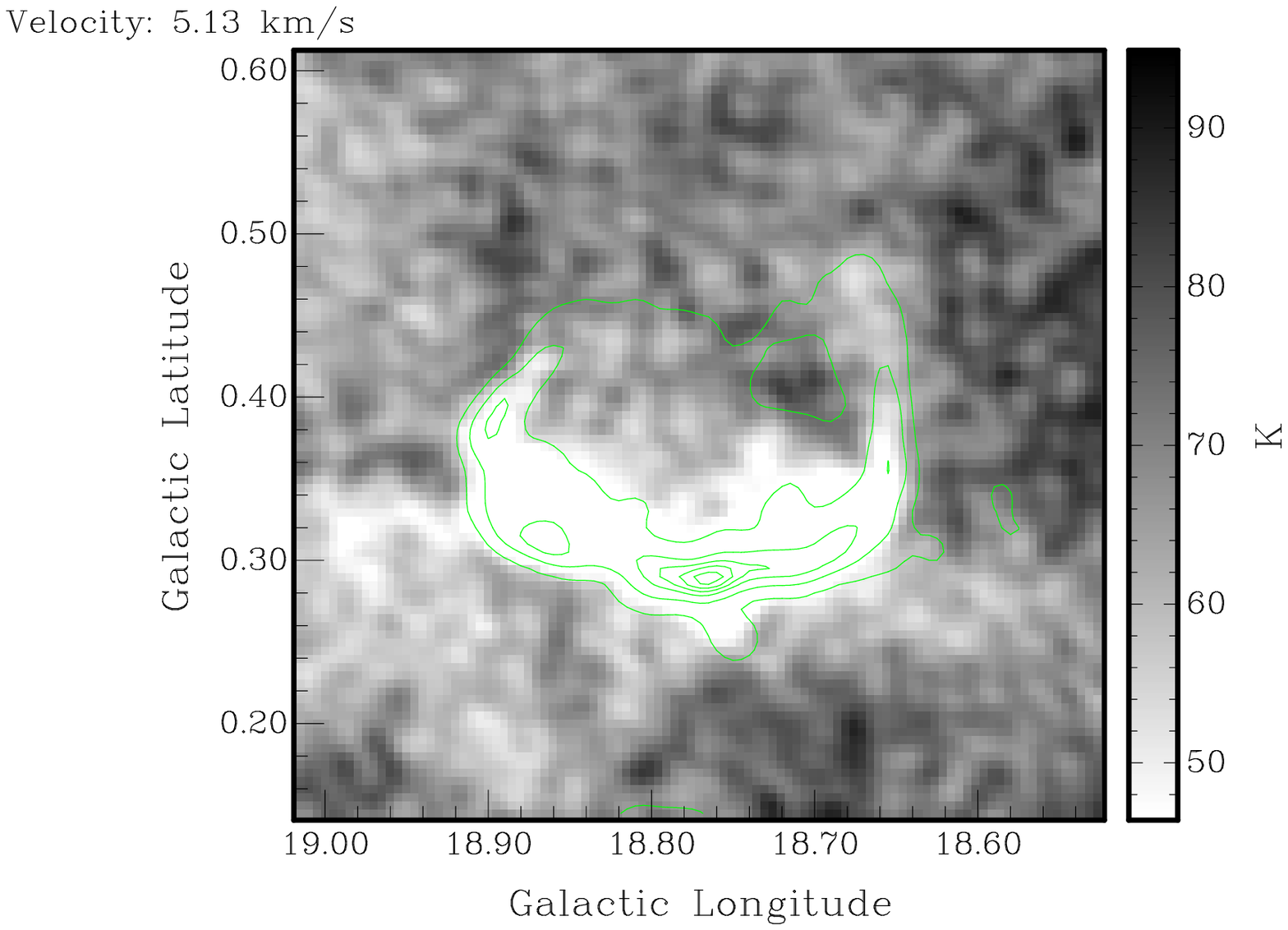}}
\put(250,110){\includegraphics{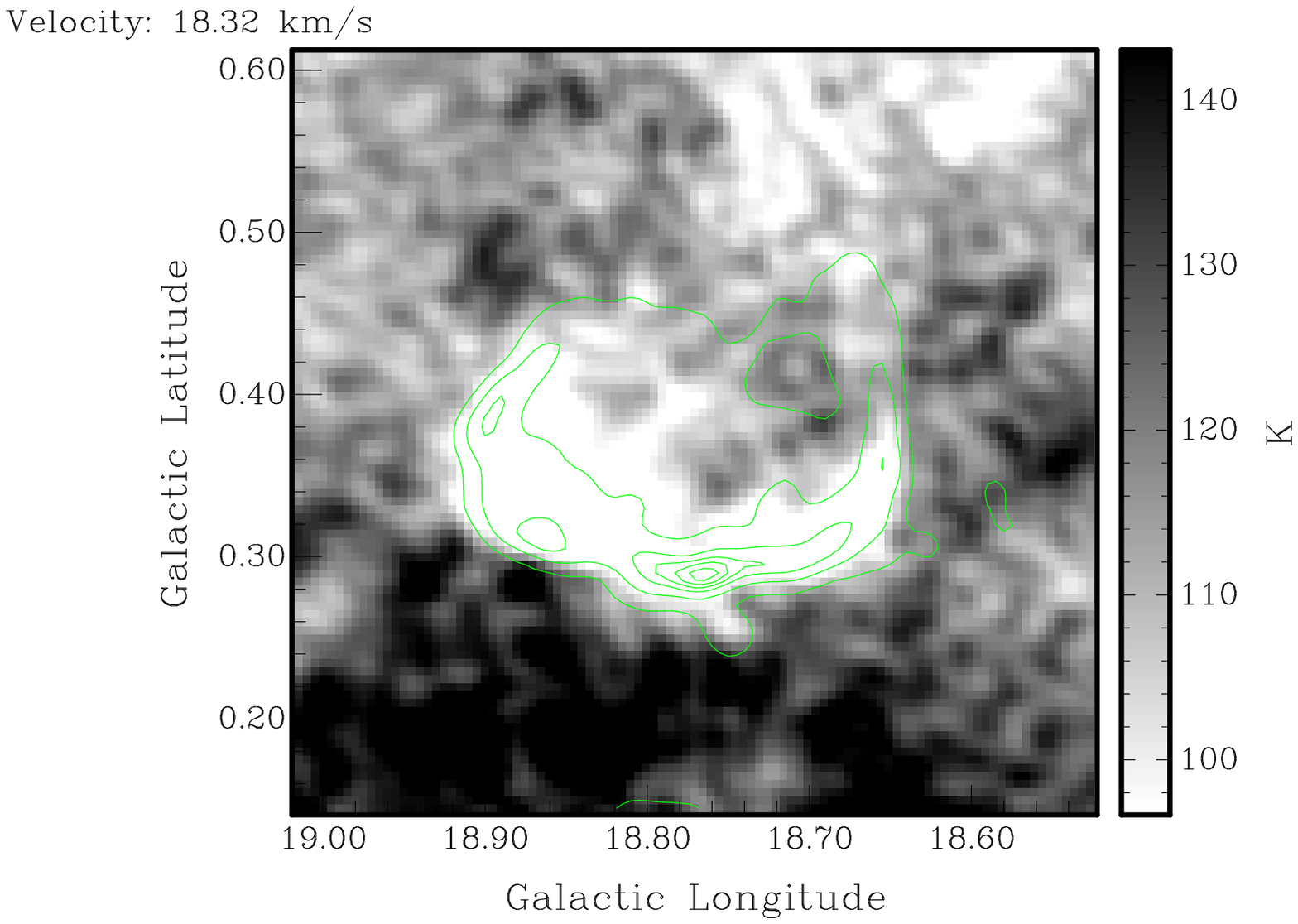}}
\put(-10,-110){\includegraphics{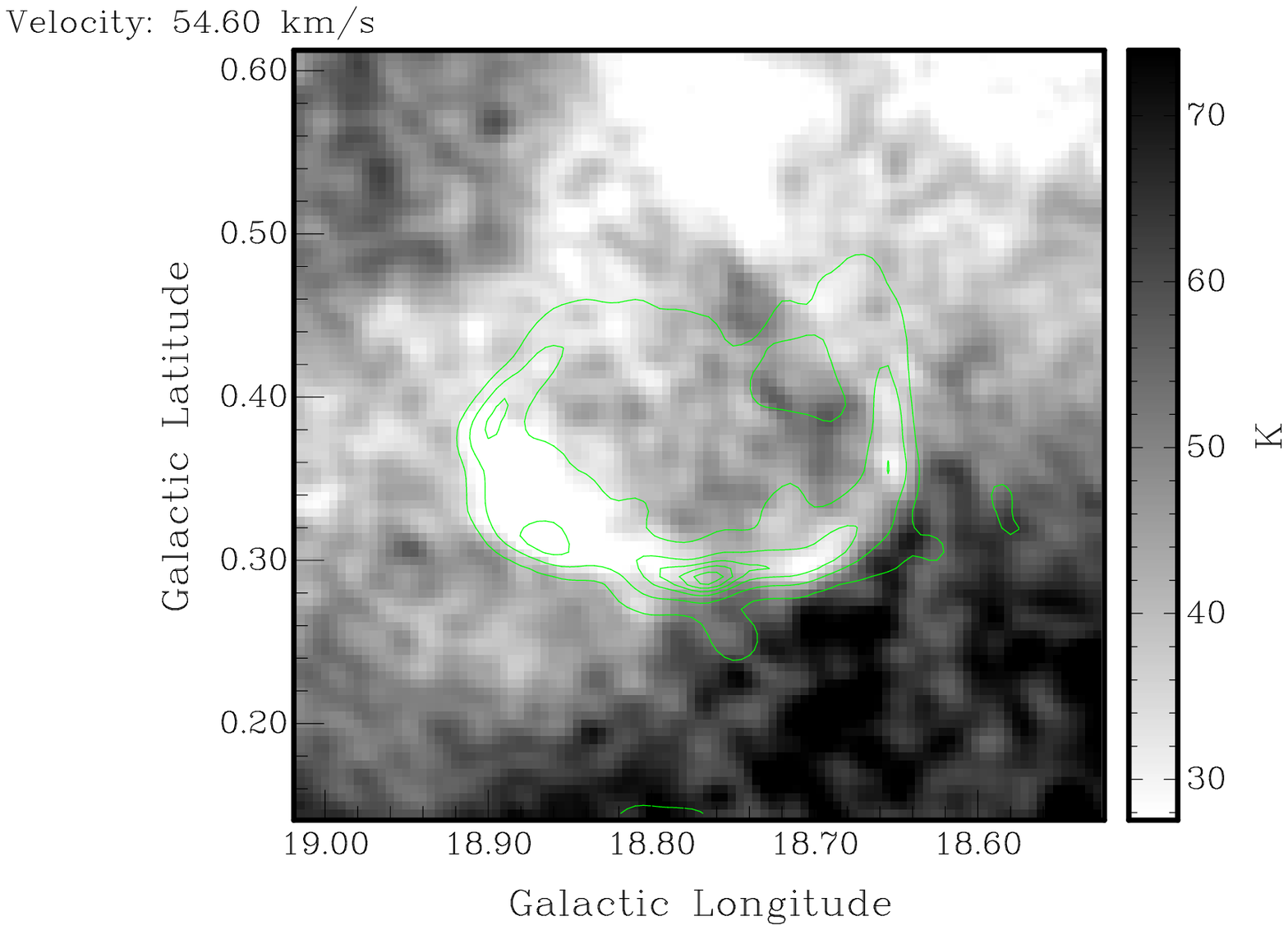}}
\put(250,-110){\includegraphics{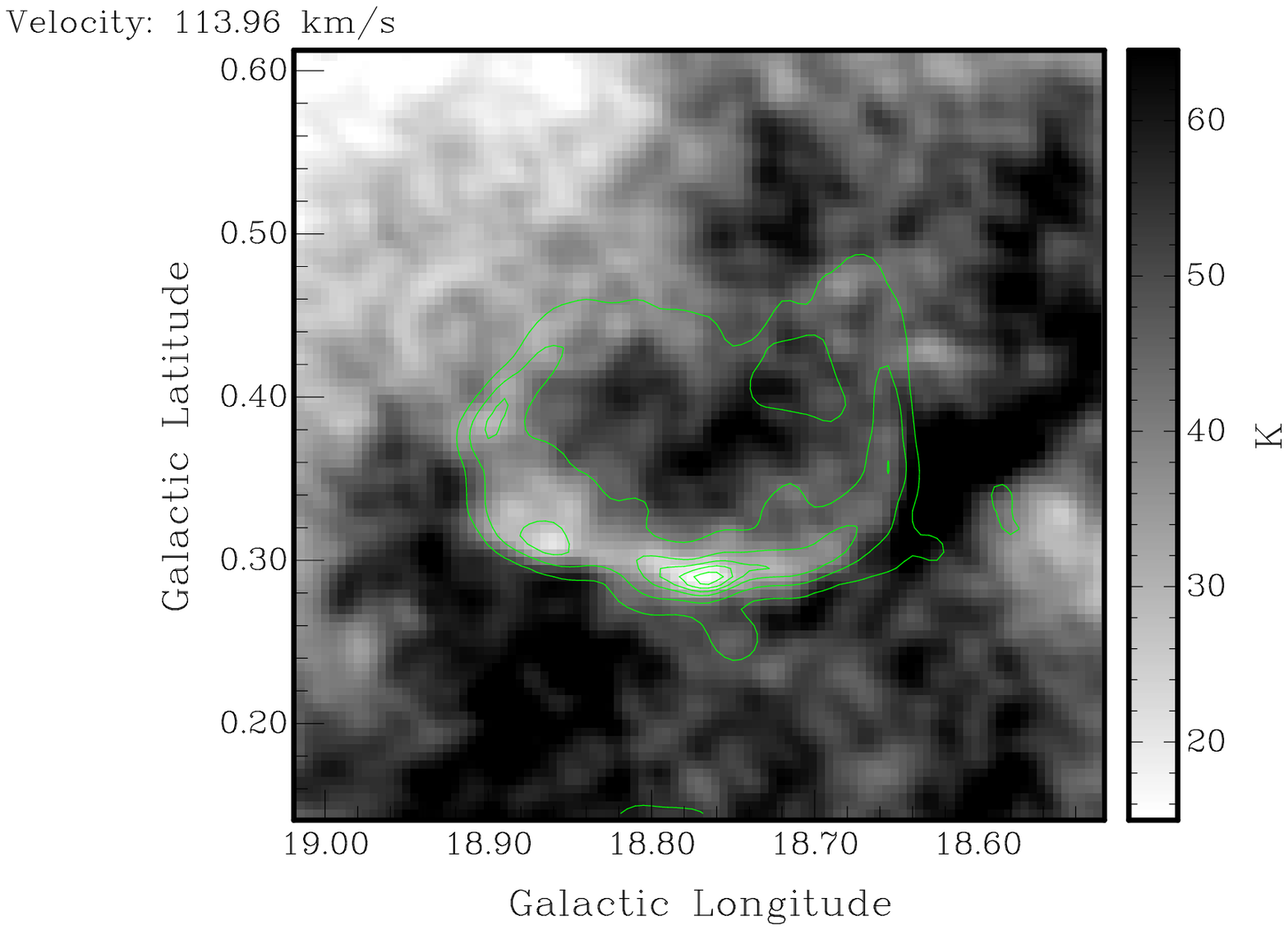}}
\end{picture}
\caption{The maps of HI emission for 4 channels: 5, 18, 54 and 114 km s$^{-1}$, respectively. These maps have superimposed contours (28, 48, 68, 80, 90, 100 K) of 1420 MHz continuum emission chosen to show the SNR.}
\end{figure*}

\subsection{Kinematic Distance}
In the HI images the highest velocity of the absorption features seen against the continuum emission of G18.8+0.3 is at 129 km s$^{-1}$.
Taking a circular Galactic rotation curve model and the most recent estimates of the parameters for this, i.e. V$_{0}$=214$\pm$7 km s$^{-1}$ (Kothes \& Dougherty 2007) and R$_{0}$=7.6$\pm$0.3 kpc (Eisenhauer et al. 2005), we obtain a circular velocity of V$_{R}$=199$\pm$2 km s$^{-1}$ at radius of the tangent point to yield a tangent point radial velocity of $\sim$ 130 km s$^{-1}$ from the HI emission spectra in Fig. 3. 
The highest velocity absorption feature is at the tangent point implying that a distance for G18.8+0.3 beyond the tangent point is required, i.e. d$_{min}$=6.9 (i.e. 7.2$\pm$0.3) kpc. Further, because there are no significant absorption features at small negative velocities, this implies that the SNR is probably within the Solar circle, i.e. an upper distance of 15 kpc applies for G18.8+0.3. The average angular diameter of 15.5$^{\prime}$ yields a diameter of about D$_{min}$=31 pc (d$_{min}$=6.9 kpc) for the SNR.

\subsection{CO Emission}
The $^{13}$CO images clearly reveal large molecular clouds that overlap the SNR in the plane of the sky. Fig. 4 is an average $^{13}$CO spectrum over the full velocity range in the direction of G18.8+0.3 (over the full area of G18.8+0.3). It shows two molecular components with high brightness temperature, at radial velocities 5$\pm$1 km s$^{-1}$ and 20$\pm$5 km s$^{-1}$ respectively, and three other molecular components with low brightness temperature, at radial velocities 54$\pm$2 km s$^{-1}$, 112$\pm$3 km s$^{-1}$ and 125$\pm$4 km s$^{-1}$ respectively. Fig. 5 (upper left) is the high-resolution $^{13}$CO image of G18.8+0.3 from a single channel at 19.65 km s$^{-1}$, which has been used to estimate G18.8+0.3's kinematic distance by previous authors. Three other plots of Fig. 5 (i.e., upper right, lower left and right) are $^{13}$CO spectra extracted from boxes 1, 2 and 3 marked in the upper left plot, respectively. These show that molecular clouds, centered at radial velocities 5 and 20 km s$^{-1}$ respectively, overlap the lower latitude part of the SNR (i.e. strong continuum emission regions).

\subsection{X-ray Emission}
Fig. 6 shows the combined X-ray image of G18.8+0.3 in the hard band (0.5-2.4 keV), with superimposed contours of the 1420 MHz continuum emission and $^{13}$CO emission at 19.65 km/s. The image consists of several detected point sources plus diffuse emission associated with the left half of the radio shell. The image in the soft band (0.1-0.3 keV) is uniform across the field of view, consistent with no emission from G18.8+0.3. A peak near l=19, b=0.4 is due to the presence of a relatively bright point-like source. The bulk, if not all, of the diffuse emission is enclosed within the radio boundary of the SNR. The apparent positional coincidence of the X-ray emission with the radio enhancement suggests that they are physically associated. With the limited spatial resolution, bandpass and sensitivity of the PSPC data, however, we cannot completely rule out the possibility for a chance line-of-sight coincidence and determine the nature of the X-ray emission. Under the assumption of a physical association and the
consideration of the SNR age, we conclude that the emission is most likely thermal,
which depends strongly on the density of hot gas. The lack of significant emission from the interior of the SNR indicates that it has a shell-like X-ray morphology, similar to that observed in radio. 
There is no X-ray emission from the right side of the radio shell which is likely associated with the presence of the CO cloud detected around 20$\pm$5 km/s. This morphological relation can be explained if the X-ray emission is produced in the entire radio shell of G18.8+0.3 (e.g. like the Cygnus Loop), but the high column density ($\sim$10$^{22}$ cm$^{-2}$, see section 5.2 for details) molecular cloud at 20$\pm$5 km/s overlapping the right side of G18.8+0.3 has completely absorbed the X-rays. There is no bright CO cloud overlapping the left side, so the diffuse X-rays are visible there. An alternative explanation is that the SNR shock has been sufficiently decelerated by interaction with the dense molecular cloud on the right side that it does not emit X-rays.

\section{Discussion}  
\subsection{Comparison of HI Absorption and CO Emission Spectra}

By comparison of CO emission and HI absorption spectra, all prominent CO molecular clouds have corresponding HI absorption, implying that all these clouds should be in front of G18.8+0.3.  Based on the Cordes $\&$ Lazio (2002) Galactic arms scheme, we conclude that our line-of-sight in the direction to G18.8+0.3 crosses the spiral arms in 7 locations. In Table 2 we list the respective spiral-arm radial velocities. Possible associations with CO emission or HI absorption features are given. Cases are included even when the velocities are up to $\sim$ 10 km s$^{-1}$ different, since the calculated arm crossing velocities are based on a circular rotation model and can be off by $\sim$ 10 km s$^{-1}$ due to non-circular motions from spiral arm shocks and velocity dispersion. 

\begin{table*}
\begin{center}
\caption{Summary of Velocity Features. }
\setlength{\tabcolsep}{1mm}
\begin{tabular}{ccccccccc}
\hline
Crossing Number:  &  local & 1 & 2 & 3 & 4 & 5  & 6 & 7 \\
\hline
 Velocity (kms$^{-1}$) & 0 & 21 & 63 & 92 & 79 & 62 & 21& 0 \\
 Nearby HI absorption feature & 0 - 5& 15 - 25 & 48 - 52 & 95 - 100& 70 - 80& 48 - 52 & 15 - 25& 0 - 5\\
 Nearby CO emission feature & 5 & 15 - 25  & 52 - 56 & 84 - 90 & 84 - 90& 52 - 56 & 15 - 25& 5\\

\hline
\hline
\end{tabular}
\end{center}
\end{table*}

Arm crossing number 4 is at the far side of the tangent point and at $v$$\approx$ 79 km s$^{-1}$. There is a significant HI absorption feature at 70 - 80 kms$^{-1}$. If this feature is associated with this spiral arm, then the absorption occurs at a distance of $\sim$ 9.4 kpc, indicating that G18.8+0.3 is beyond this distance.  We note that the absorption feature for arm crossings 5, 6 and 7, all lying beyond the tangent point, occur at the same velocities as for arm crossings 2, 1 and the local gas. Thus, the HI absorption or CO emission features at these velocities do not provide evidence on whether G18.8+0.3 is beyond crossings 5, 6 and 7. Two features seen in both the HI absorption and CO emission spectra near the tangent point (i.e., at velocities of 110 - 115 and 122 - 130 km s$^{-1}$) are not noted as corresponding a spiral arm in the Cordes $\&$ Lazio diagram. We identify them as probably originating from the galactic molecular ring.

\begin{figure*}
\vspace{200mm}
\begin{picture}(80,80)
\put(50,400){\includegraphics{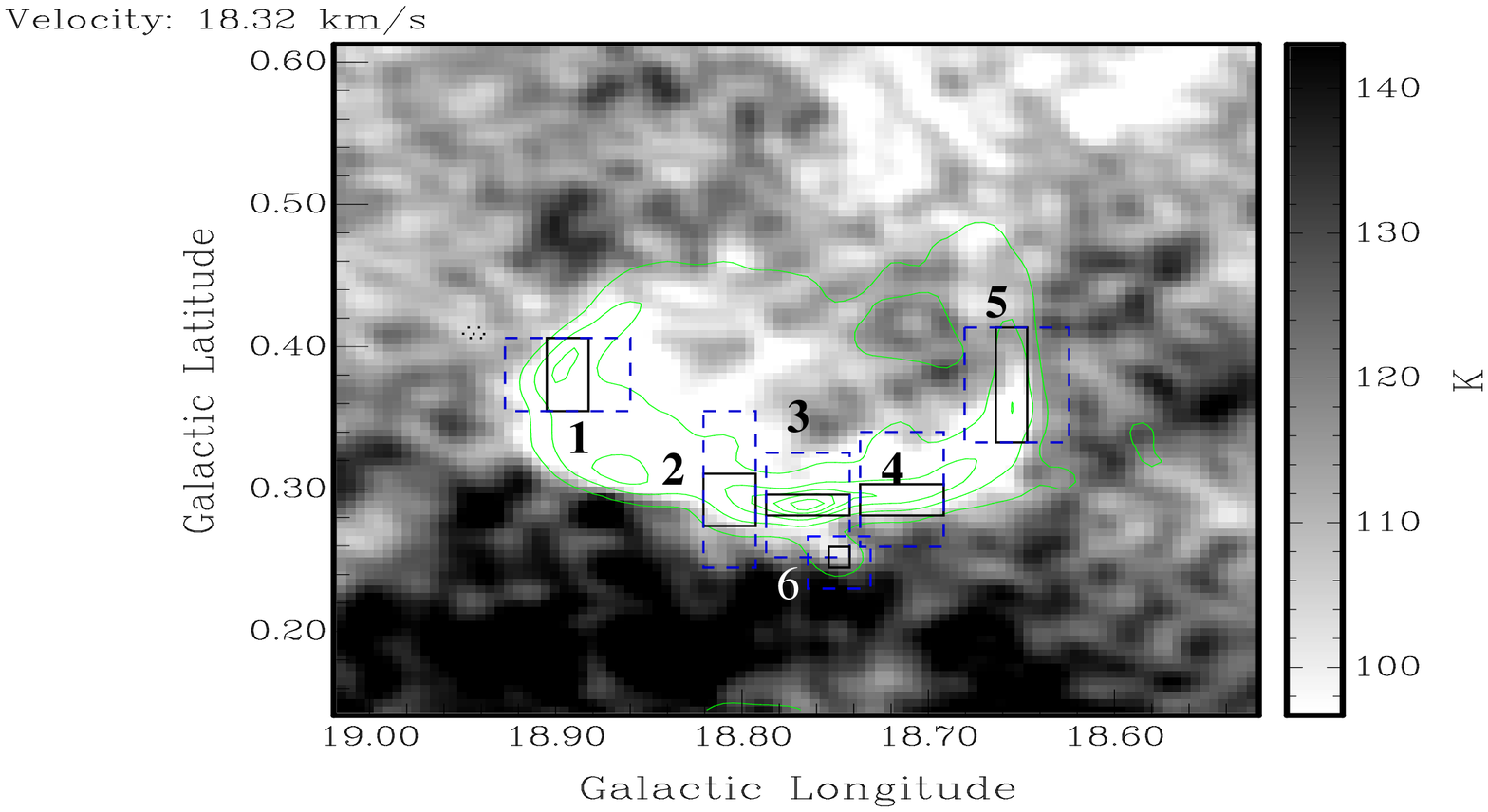}}
\put(-10,285){\includegraphics{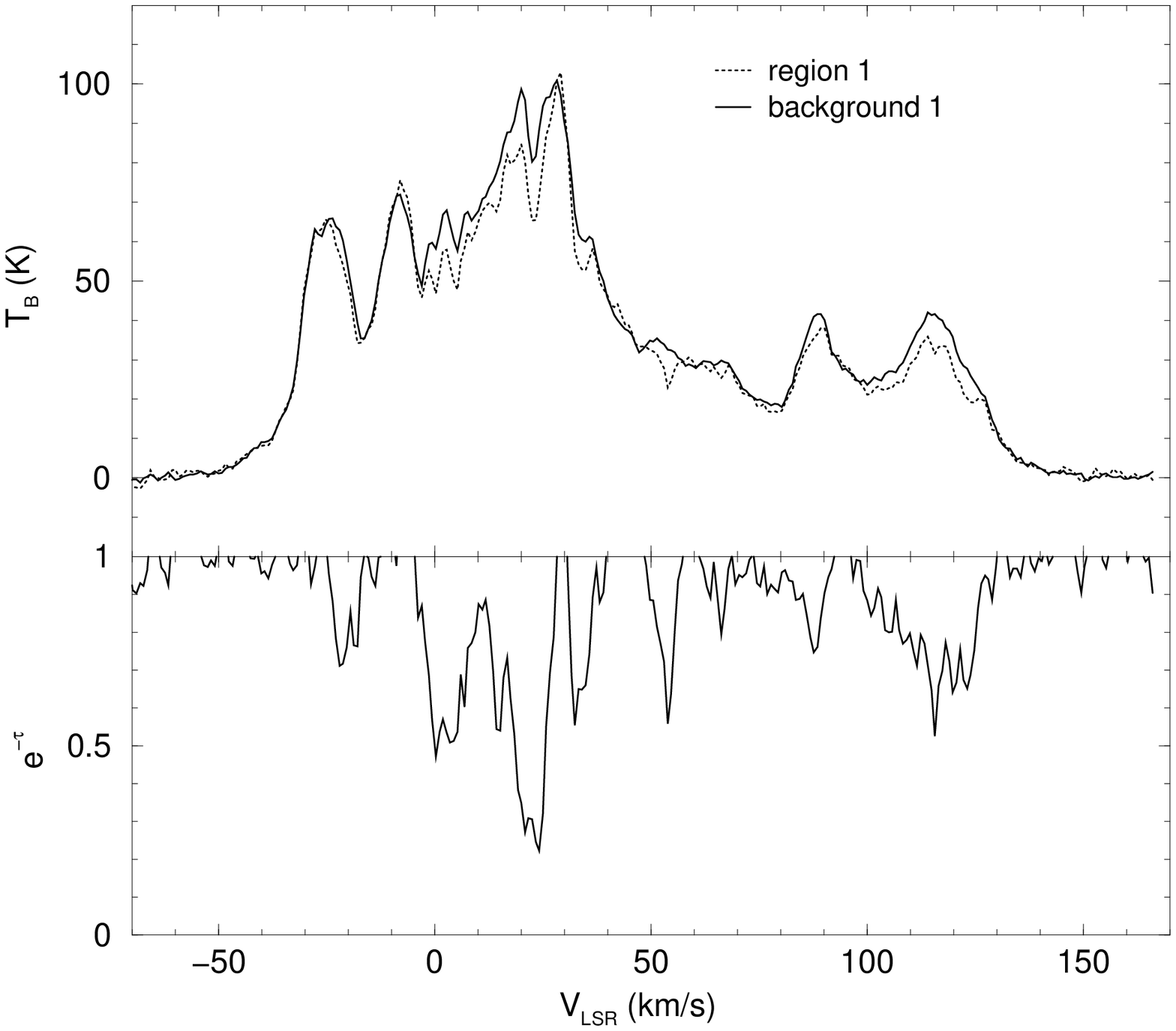}}
\put(270,285){\includegraphics{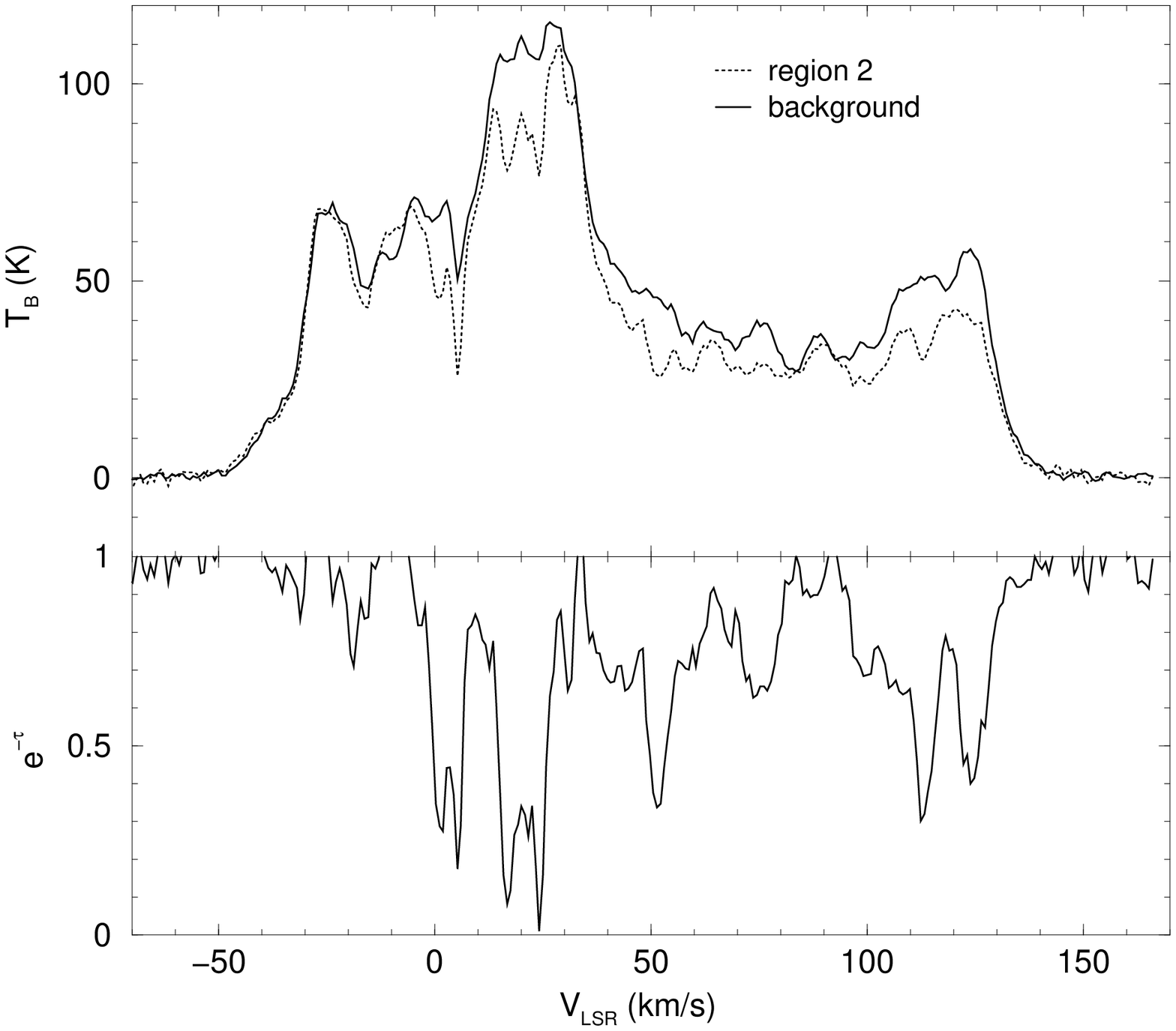}}
\put(-10,135){\includegraphics{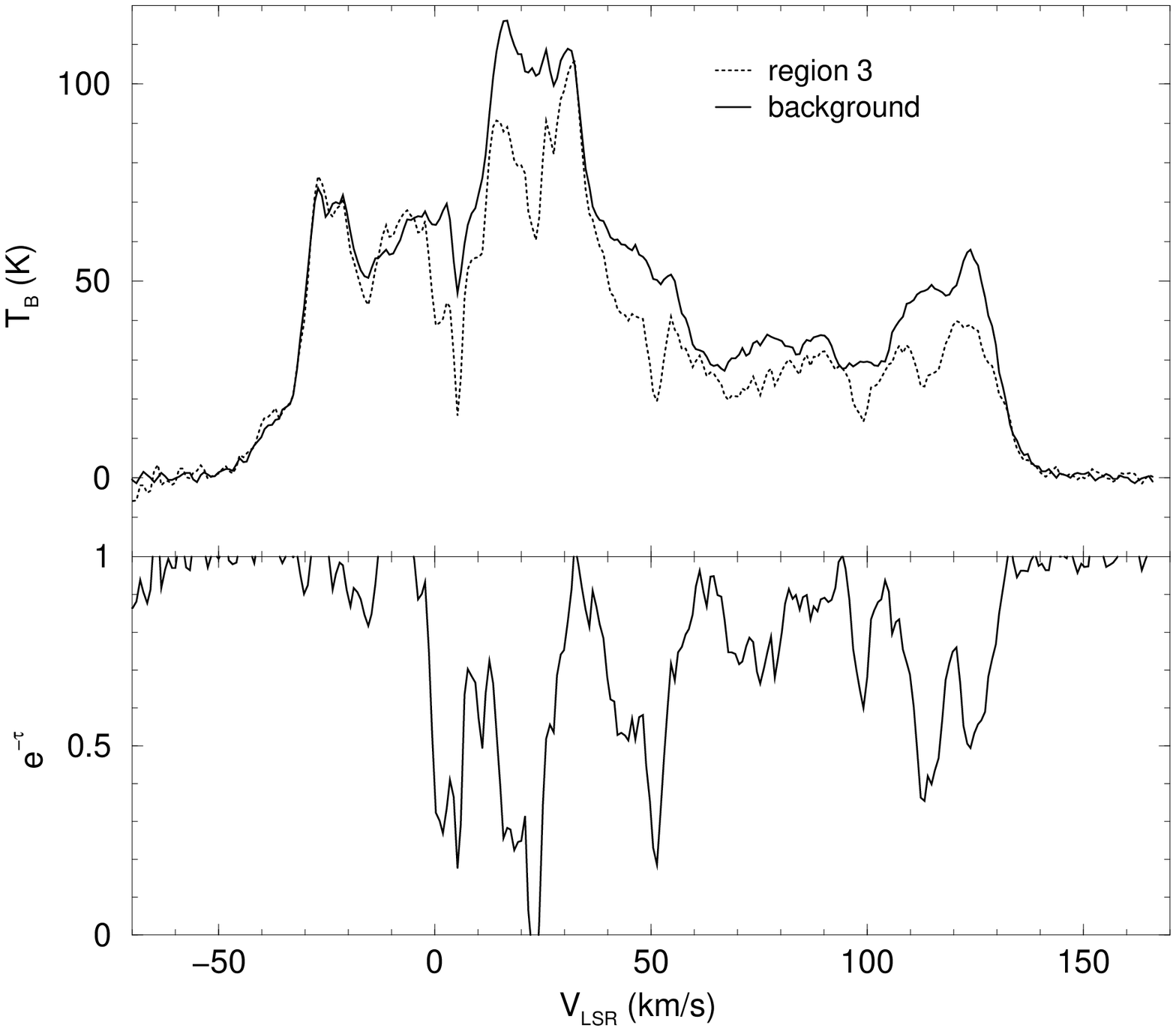}}
\put(270,135){\includegraphics{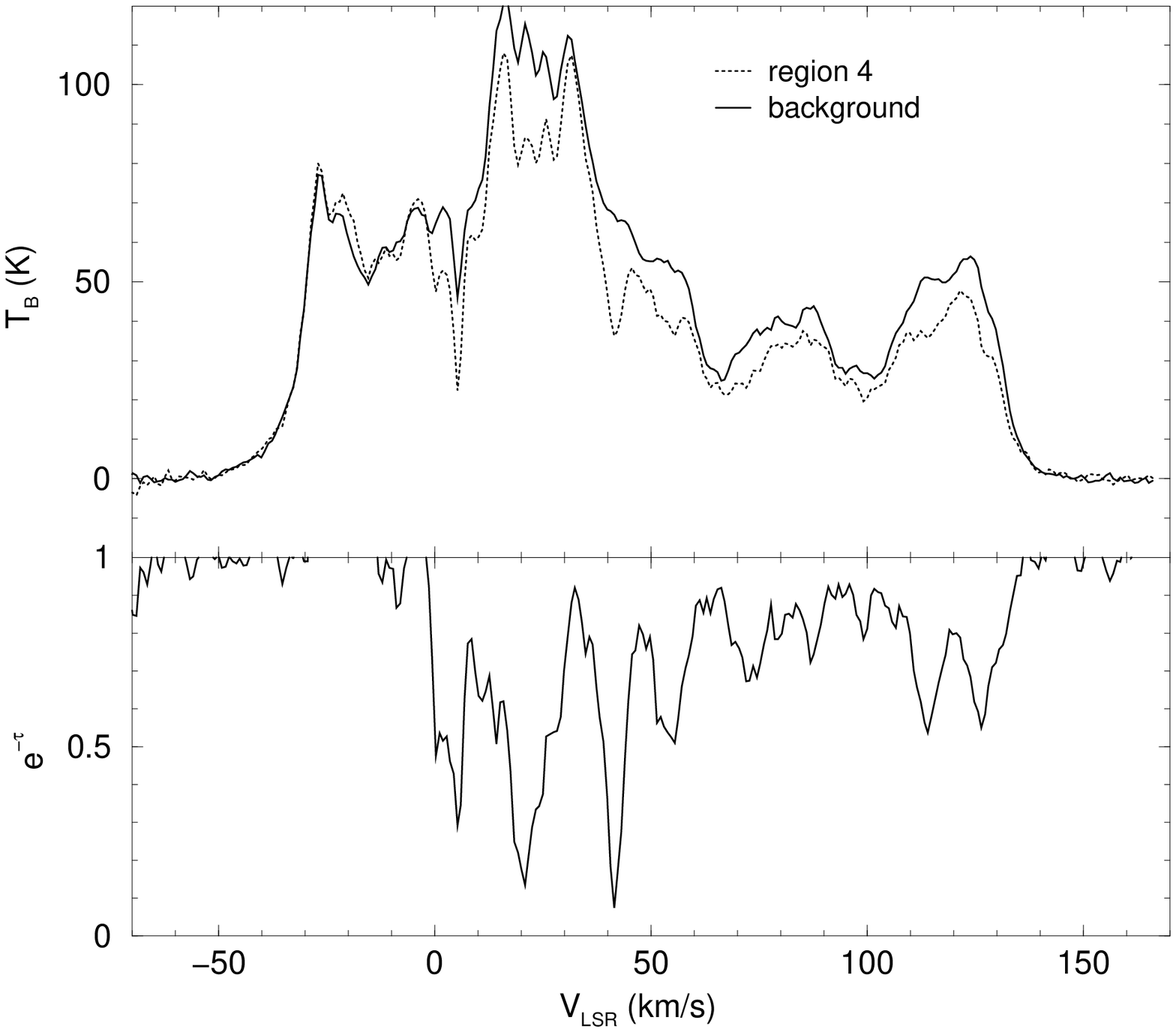}}
\put(-10,-15){\includegraphics{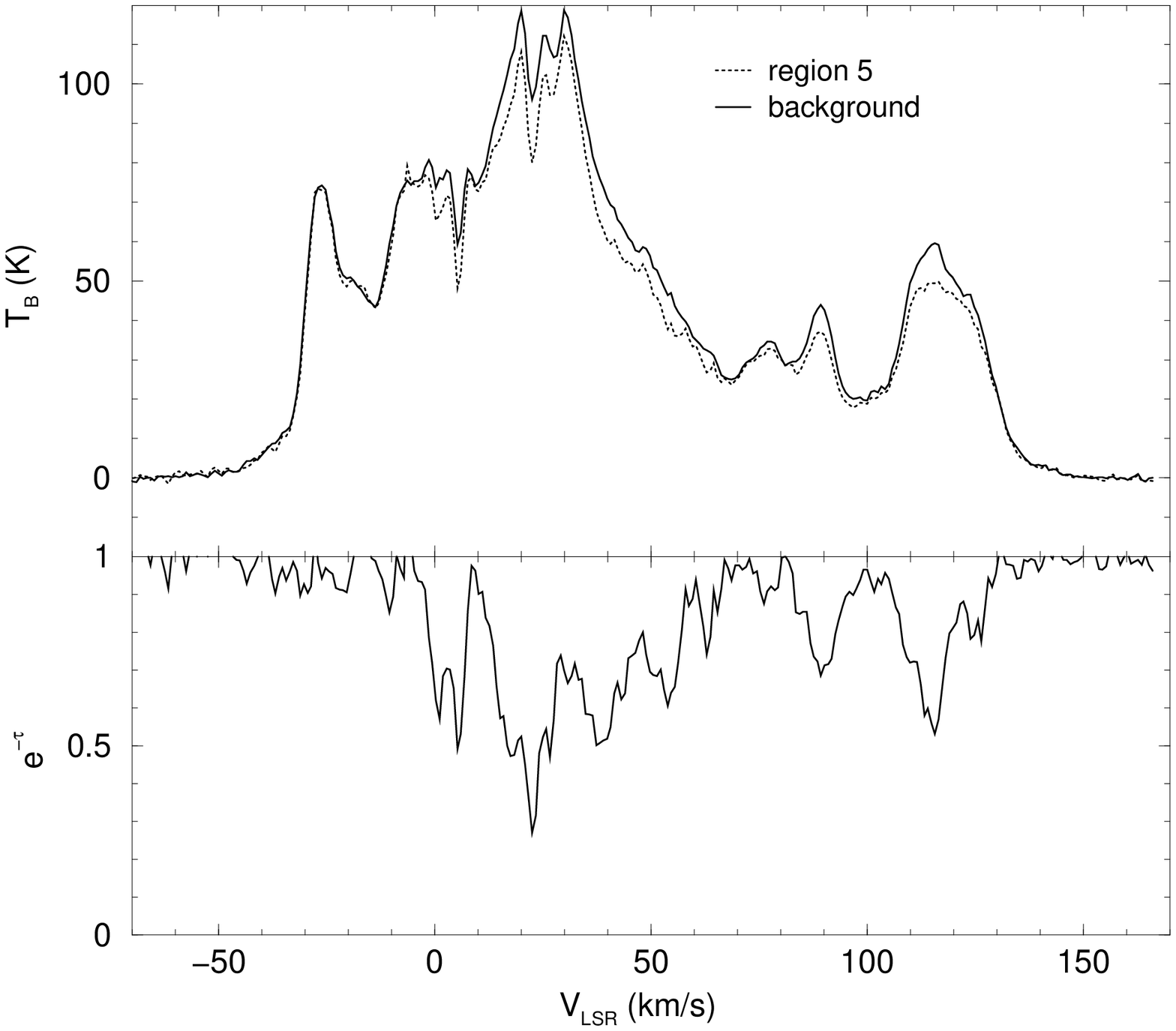}}
\put(270,-15){\includegraphics{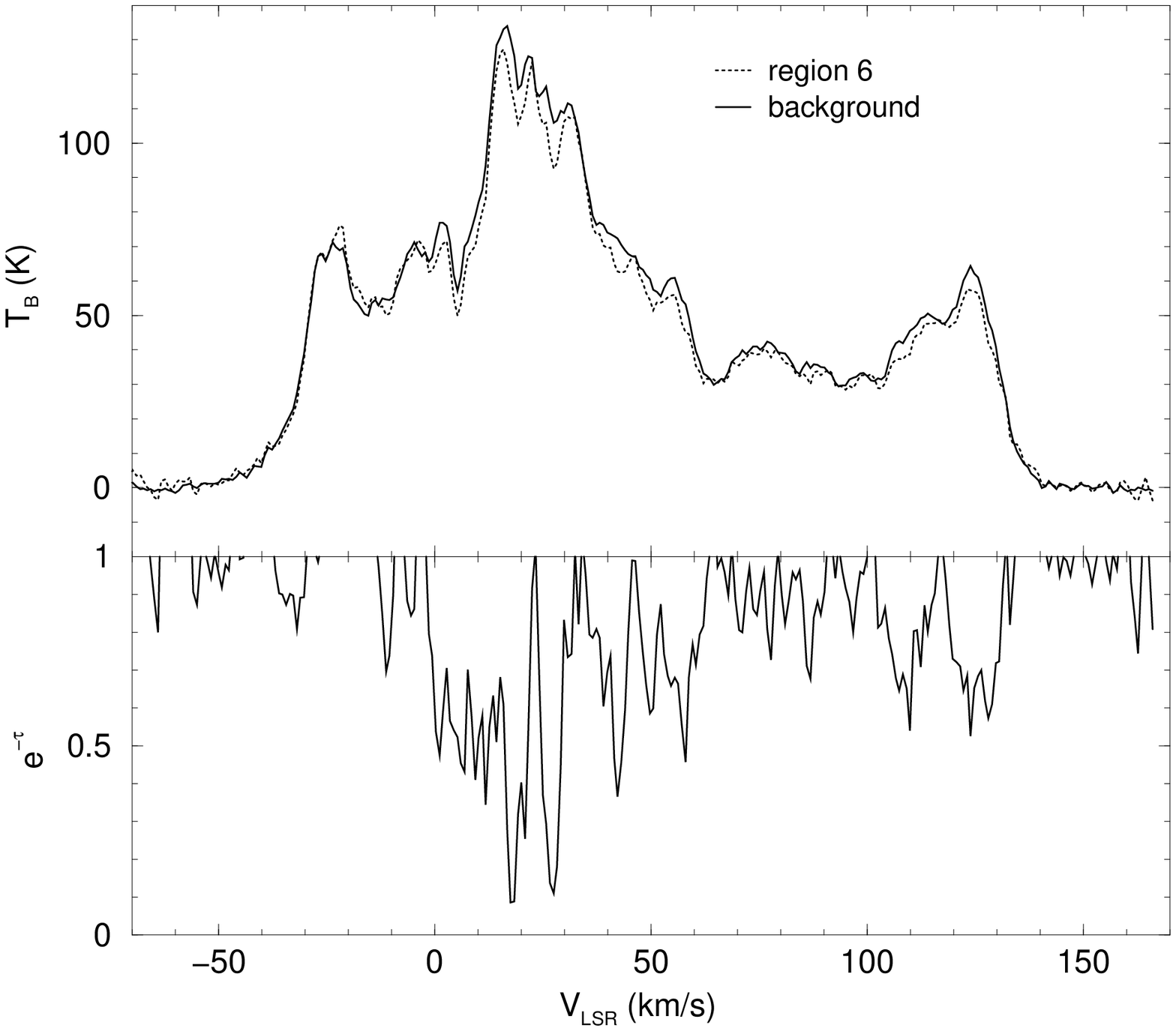}}
\end{picture}
\caption{HI image of G18.8+0.3 from a single channel at 18.3 km s$^{-1}$ (top row). The six other panels show the HI spectra, extracted from boxes 1, 2, 3, 4, 5 and 6 shown in the first panel. The HI emission and absorption spectra clearly show absorption features at 1, 5, 11, 24, 27, 51, 98, 113, 126 km s$^{-1}$.}
\end{figure*}

\subsection{Is G18.8+0.3 Interacting with a Dense CO Cloud?}
Dubner et al. (1999, 2004), mainly based on their moderate-resolution $^{12}$CO line and low-resolution HI line observations, suggested a physical association between G18.8+0.3 and the GMC detected by the $^{12}$CO observations at around 19 km s$^{-1}$. They estimated a distance of 14 kpc for G18.8+0.3 after considering the low-resolution HI absorption profiles from Caswell et al. (1975). However, they found no detection of significant broadening in the individual $^{12}$CO spectra,  while a broadened profile would be considered as a clear indicator of shock interaction.

Fig. 4 reveals that the molecular cloud at 20$\pm$5 km s$^{-1}$ is the only one in the $^{13}$CO spectra to show a clearly broadened profile. The high resolution image of $^{13}$CO (see, Fig. 5) shows the morphological agreement between the shape of this $^{13}$CO cloud and the border of the SNR more clearly than does that of Dubner et al. (1999). This detailed coincidence can be interpreted as an interaction between the SNR shock and the CO cloud. 
If true, this CO line feature (or perhaps only part of it) is due to a cloud at the far kinematic distance at 20 kms$^{-1}$ (i.e. 12.1 kpc) which would also be the distance independently estimated for G18.8+0.3. The total $H_{2}$ mass (upper limit) of the CO cloud at 20$\pm$5 km s$^{-1}$ is estimated from $M_{H_{2}}$ = $N_{H_{2}}$$\Omega$$\it{d}^{2}$(2m$_{H}$/$M_{\odot}$) ($\Omega$ is the solid angle of the cloud, and $\it{d}$ is its distance). We take $N_{H_{2}}$/$W_{^{12}CO}$$\approx$1.8$\times$10$^{20}$ cm$^{-2}$ K$^{-1}$ km$^{-1}$s from Dame et al. (2001). The total integrated intensity of $^{13}$CO is $W_{^{13}CO}$ $\approx$ 2.5 K km/s from Fig. 4 (over a 0.2$^{o}$$\times$0.3$^{o}$ region). Because the $^{12}$CO/$^{13}$CO isotopic abundance ratio is about 70 at this distance (Langer \& Penzias 1990), we obtain an average $H_{2}$ column density of $N_{H_{2}}$$\approx$3.2$\times$10$^{22}$ cm$^{-2}$, and a molecular cloud mass of $M_{H_{2}}$$\approx$6$\times$$10^{5}$$M_{\odot}$. This giant molecular cloud has a mean density of $\sim$ 2$\times$ 10$^{2}$ cm$^{-3}$.

Since the CO should be in front of G18.8+0.3, as implied by the strong HI absorption feature detected near $v$$\approx$20 km$s^{-1}$, it may be adjacent to or surrounding the SNR. Shock-excited OH maser emission at 1720 MHz may result from the interaction of a SNR with an adjacent molecular cloud, but this kind of OH maser is only efficiently pumped in hot, dense shocked molecular clouds under very restrictive conditions (Wardle 1999; Wardle \& Yusef-Zadeh 2002). The failure to detect OH masers by Dubner et al. (1999) could mean that either that the angle at which the shock propagates transverse to the line-of-sight is small, that the shocked molecular cloud is not sufficiently dense (e.g. above 10$^{4}$cm$^{-3}$), or that the OH column density is not high enough (e.g. above 10$^{16}$ cm$^{-2}$) to produce 1720 MHz masers.
 
\begin{figure} 
\vspace{50mm}
\begin{picture}(55,55)
\put(-10,200){\includegraphics{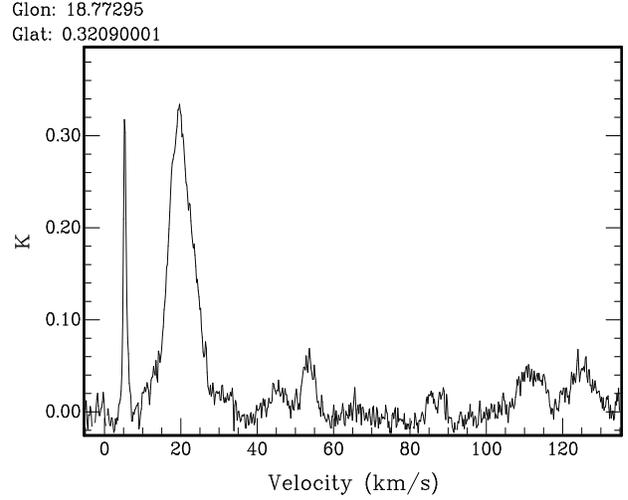}}
\end{picture}
\caption{The average $^{13}$CO spectrum over the full velocity range in the direction of G18.8+0.3.} 
\end{figure} 

\begin{figure*}
\vspace{120mm}
\begin{picture}(80,80)
\put(0,185){\includegraphics{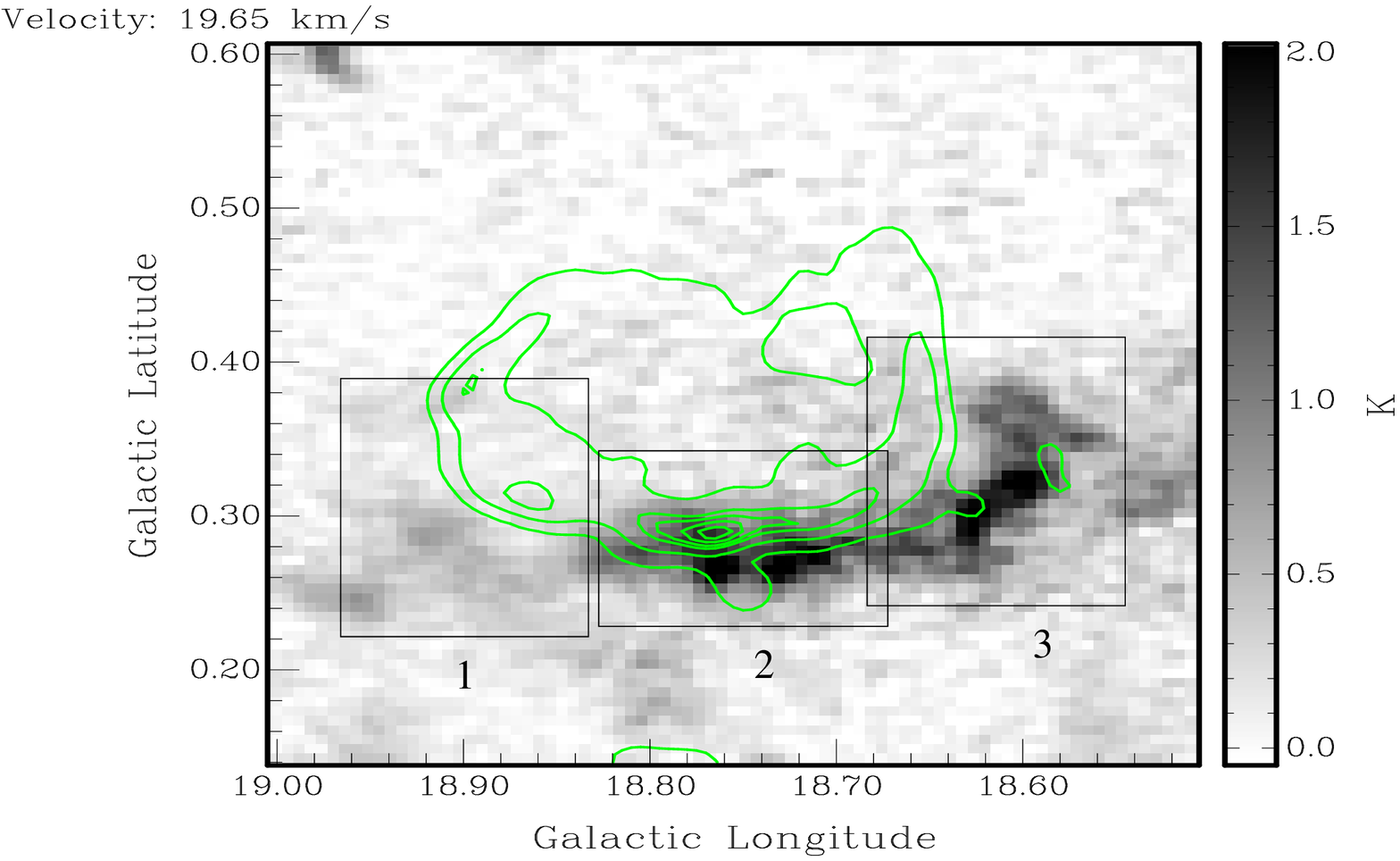}}
\put(270,430){\includegraphics{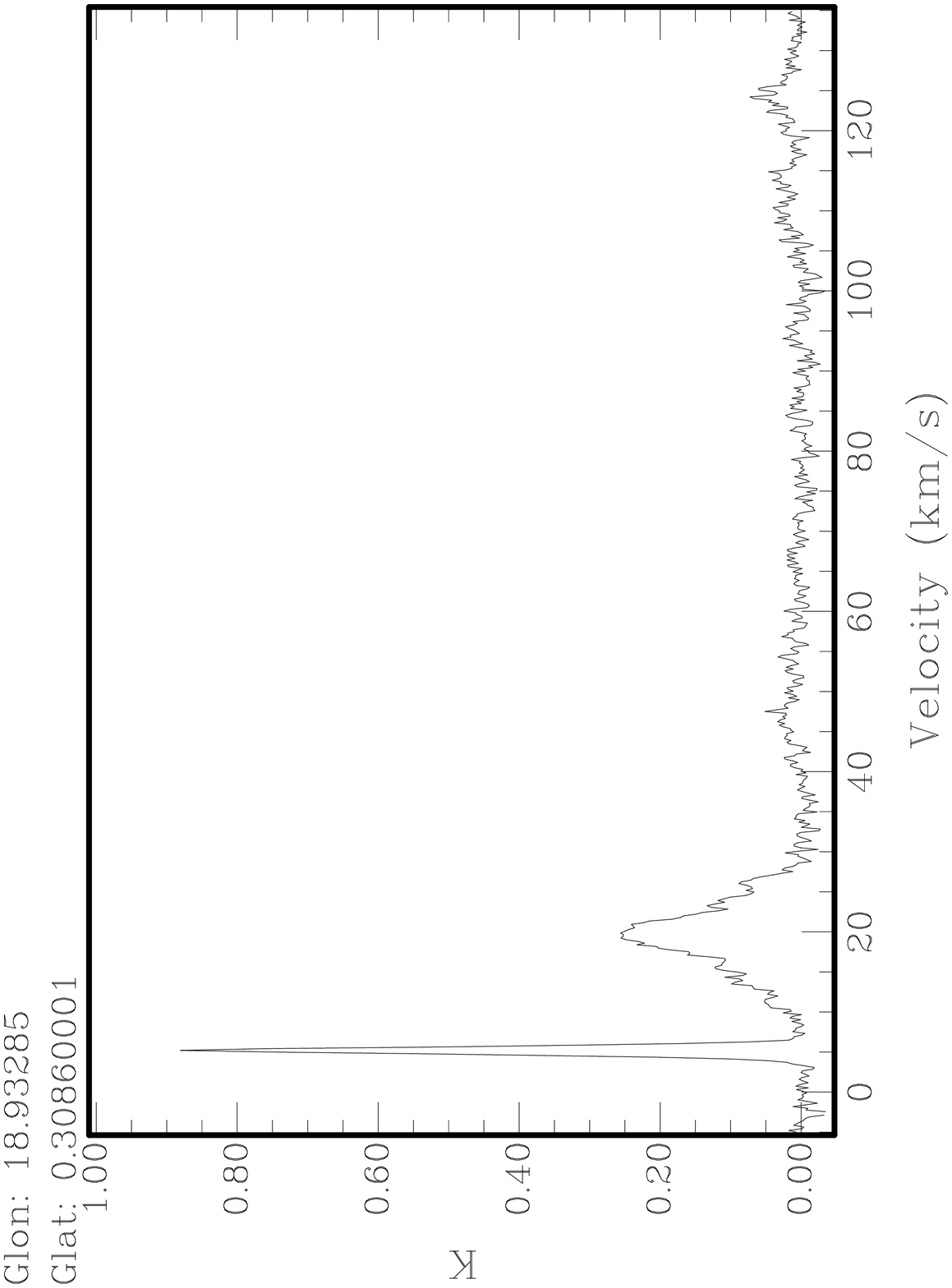}}
\put(5,210){\includegraphics{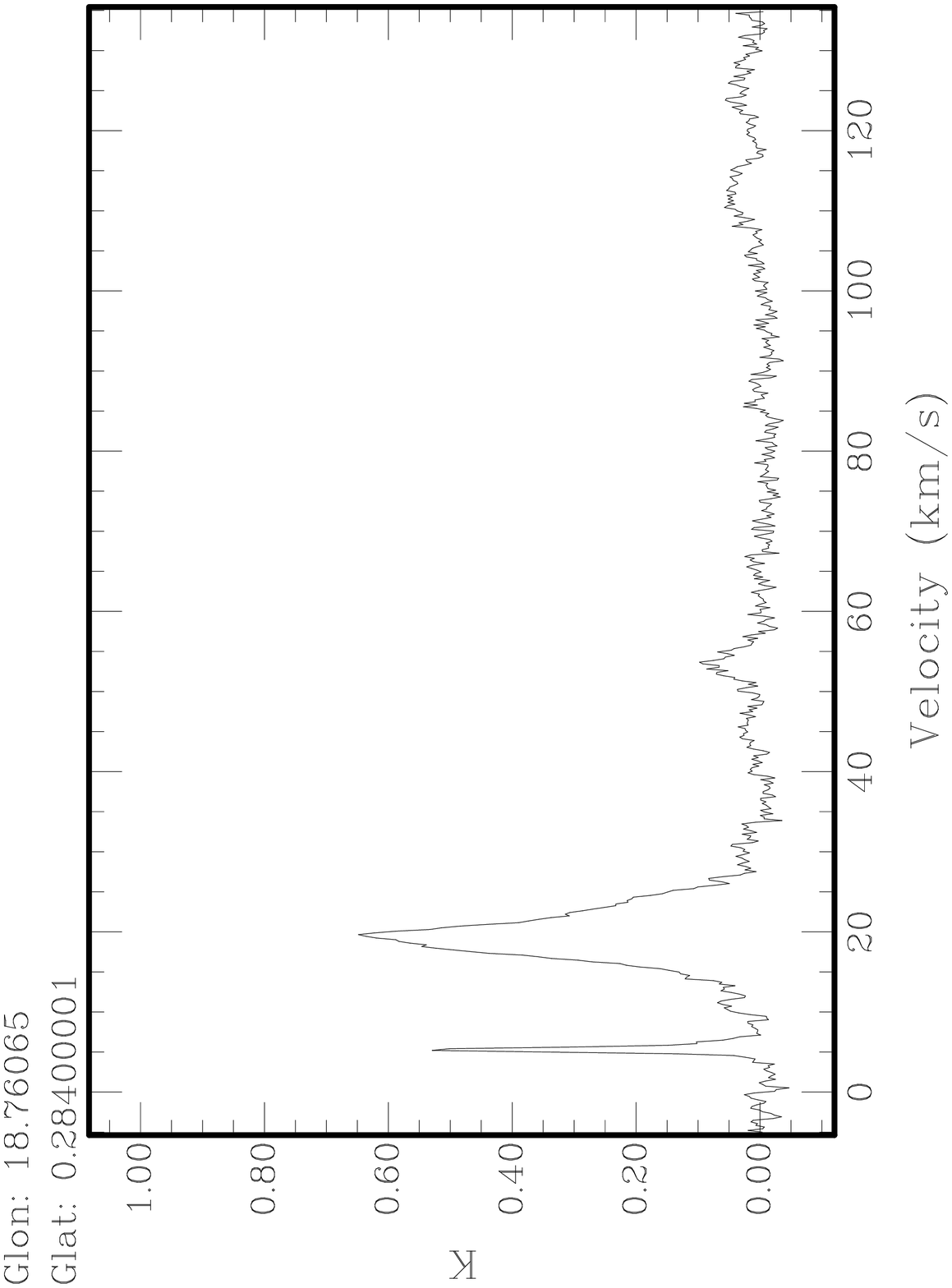}}
\put(270,210){\includegraphics{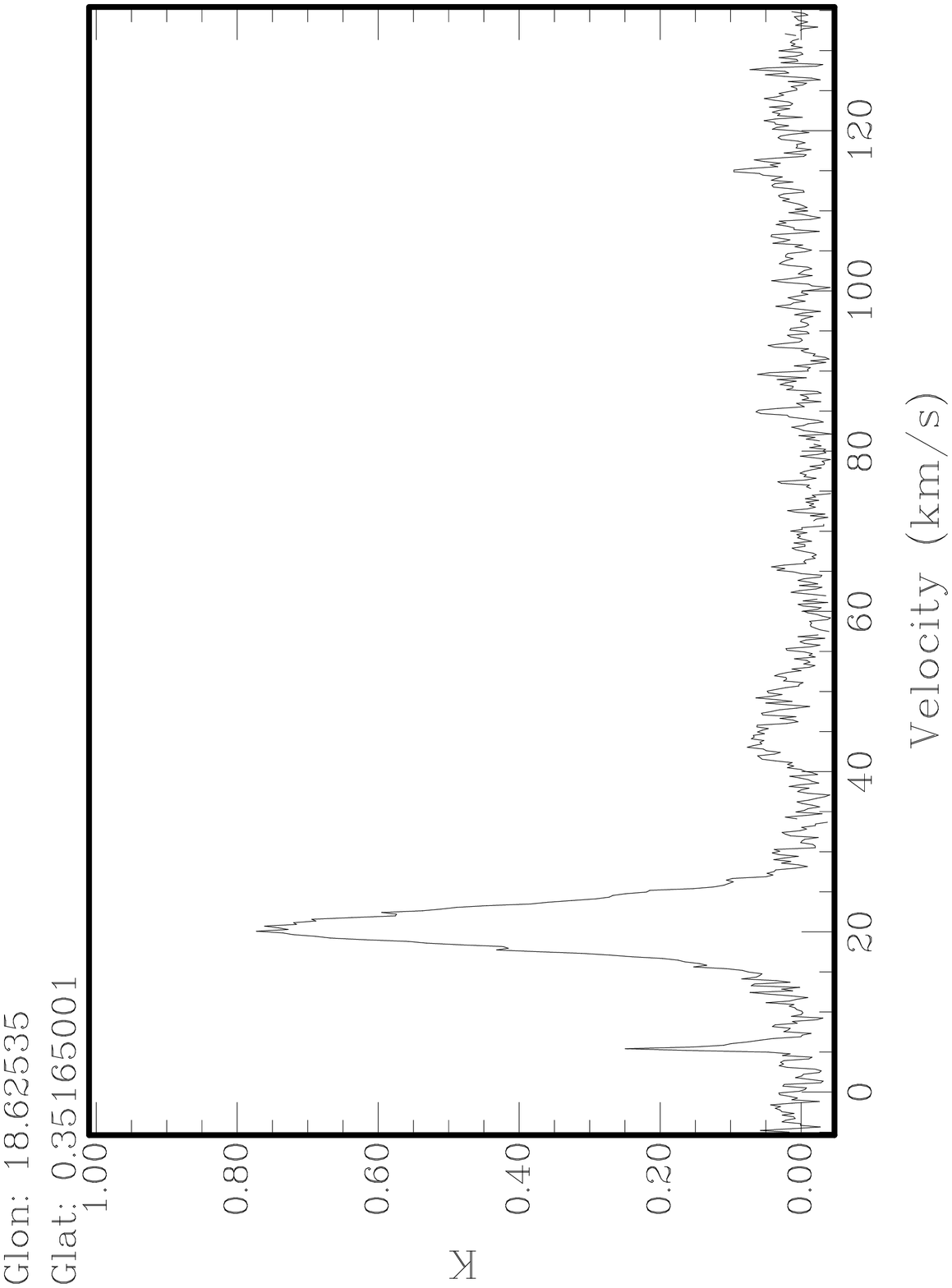}}
\end{picture}
\caption{ The upper left of Fig. 5 is the high-resolution $^{13}$CO image of G18.8+0.3 from a single channel at 19.84 km s$^{-1}$. The spectra in three other plots, i.e., the upper right, the lower left and right, are extracted from the boxes 1, 2 and 3 marked in the upper left plot, respectively. }
\end{figure*}

\subsection{X-ray Luminosity}
As demonstrated in Section 4.2, a likely interaction between the SNR and an adjacent molecular cloud implying their physical association. The spatial correlation between X-ray and radio emission noted in Section 3.5 suggests that the origin of the X-ray emission is related to the SNR shock, i.e. is due to bremsstrahlung from the shock-heated gas. 

Assuming the gas temperature $T_{s}$, we can obtain the absorbing column density $N_{HI}$ from the HI absorption spectrum ($N_{HI}$=1.9$\times$10$^{18}$$\tau$$\Delta{\it{v}}$$T_{s}$ cm$^{-2}$, Dickey \& Lockman 1990). Using the region 3 spectra shown in Fig. 3, we sum over the absorption features to obtain $N_{HI}$ $\sim$2$\times$10$^{22}$ cm$^{-2}$, taking a typical value of $T_{s}$= 100 K. 

Based on plasma emission models, the present X-ray data can be used to estimate the X-ray flux and intrinsic luminosity (at least an upper limit) of G18.8+0.3.  From the X-ray emitting area (0.15$^{o}$$\times$0.20$^{o}$) seen against the face of G18.8+0.3 in the Fig. 6, a count rate of 7.3$\times$10$^{-3}$ counts s$^{-1}$ is obtained. To get the intensity of the diffuse X-ray emission associated with the remnant, we first obtain a total count rate within this region above a local background and then subtract the contribution from the four X-ray point sources detected within the region.
Assuming a Raymond Smith plasma emission model, using W3pimms (http://heasarc.gsfc.nasa.gov/Tools/w3pimms.html), and taking $N_{HI}$$\sim$2$\times$10$^{22}$ cm$^{-2}$, an energy range of 0.5 - 2.4 keV, and a temperature 1 keV, the X-ray flux, unabsorbed flux and intrinsic luminosity of G18.8+0.3 are F$_{X}$=1.5$\times$10$^{-13}$ erg cm$^{-2}$ s$^{-1}$, F$^{un}_{X}$=2.2$\times$10$^{-12}$ erg cm$^{-2}$ s$^{-1}$ and L$_{X}$=3.6$\times$10$^{34}$ erg s$^{-1}$ (L$_{X}$=4$\pi$d$^{2}$F$^{un}_{X}$, where we take d=12.1 kpc), respectively.

There is no X-ray emission from the right side of the SNR which is covered by the GMC at 20$\pm$5 km/s. The GMC has an average $H_{2}$ column density of $N_{H_{2}}$$\approx$3.2$\times$10$^{22}$ cm$^{-2}$. This yields $N_{HI+H_{2}}$$\sim$5$\times$10$^{22}$ cm$^{-2}$. If the unabsorbed X-ray flux from the right side (behind the GMC) is still 2.2$\times$10$^{-12}$ erg cm$^{-2}$ s$^{-1}$ (T= 1 keV), the surface brightness after absorption is 8.3$\times$10$^{-6}$ counts $s^{-1}$ arcmin$^{-2}$ (total 6.7$\times$ 10$^{-4}$ counts s$^{-1}$). This is below the detection limit of the PSPC observations, so this lack of observed X-ray emission is consistent with foreground absorption. 

\begin{figure}
\vspace{50mm}
\begin{picture}(60,60)
\put(-130,285){\includegraphics{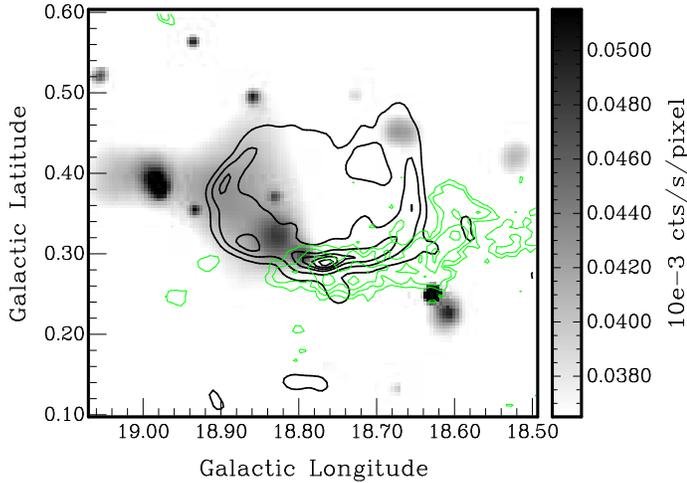}}
\end{picture}
\caption{The ROSAT PSPC intensity image of G18.8+0.3 in the hard band (0.5-2.4 keV). The intensity is adaptively smoothed using csmooth to achieve a signal-to-noise ratio of $\sim$ 3. The image has superimposed on contours of $^{13}$CO emission (green line: 0.6, 0.9, 1.4, 2.1 K) and 1420 MHz continuum emission as Fig. 2. The units of the intensity are 10$^{-3}$ counts s$^{-1}$ pixel $^{-1}$ (1 pixel=0.25$^{\prime}\times 0.25^{\prime}$)}
\end{figure}

\section{Conclusion}
Summing up, using new VLA continuum and HI-line data for G18.8+0.3, we have obtained HI absorption profiles towards this SNR. From the comparison of the HI absorption and $^{13}$CO emission profiles in the direction of G18.8+0.3, we obtain a new estimate for the distance and physical size of G18.8+0.3. G18.8+0.3 shows HI absorption features up to the tangent point velocity, giving a lower distance limit of 6.9 kpc. We detect the HI absorption and CO emission coincident with the expected velocity for a spiral arm at 9.4 kpc according to the Cordes $\&$ Lazio (2002) model for Galactic arms. This suggests that G18.8+0.3 is probably located beyond a distance of 9.4 kpc. Absence of HI absorption at negative velocities gives an upper distance limit of 15 kpc for the SNR. Morphological agreement between G18.8+0.3 and a CO cloud, plus the broadened nature of the CO emission profile support an interaction between G18.8+0.3 and this molecular cloud. This implies a distance of 12.1 kpc for G18.8+0.3. ROSAT PSPC observations reveal  diffuse X-ray emission associated with the left side of the radio shell of G18.8+0.3. 
We estimate an atomic hydrogen column density of $N_{HI}$ of $\sim$ 2$\times$10$^{22}$ cm$^{-2}$, and an intrinsic X-ray flux and luminosity for G18.8+0.3.

\begin{acknowledgements}We thank the referee for constructive suggestions which helped to improve the presentation of the paper. We wish to thank Chris Salter at National Astronomy and Ionosphere Center for careful reading of the manuscript. WWT and DAL acknowledge support from the Natural Sciences and Engineering Research Council of Canada. WWT appreciates support from the Natural Science Foundation of China. This publication makes use of X-ray data from the High Energy Astrophysics Science Archive Reseacher Center (HEASARC), provided by NASA's Goddard Space Flight Center, and molecular line data from the Boston University-FCRAO Galactic Ring Survey (GRS). The GRS is a joint project of Boston University and Five College Radio Astronomy Observatory, funded by the National Science Foundation. The NRAO is a facility of the National Science Foundation operated under cooperative agreement by Associated Universities, Inc.
\end{acknowledgements}


\begin{thebibliography}{}
\bibitem[2006]{Ahaet06}Aharonian, F., Akhperjanian, A.G., Bazer-Bachi, A.R. et al. 2006, ApJ, 636, 777
\bibitem[1975]{Caset75}Caswell, J.L., Murray, R.S., Roger, R.S. et al. 1975, A\&A, 45, 239
\bibitem[2002]{Coret02}Cordes, J.M. \& T.~Joseph~W.~Lazio 2002, astro-0207156
\bibitem[1986]{Damet86}Dame, T.M., Hartmann, D. and Thaddeus, P. 2001, ApJ, 547, 792 
\bibitem[1990]{Dicet90}Dickey, J.M., Lockman, F. J. 1990, Annu. Rev. A\&A, 28, 215
\bibitem[2004]{Dubet04}Dubner, G., Giacani, E., Reynoso, E., Parón, S. 2004, A\&A, 426, 201.
\bibitem[1999]{Dubet99}Dubner, G., Giacani, E., Reynoso, E., Goss, W. M., Roth, M., Green, A. 1999, AJ, 118, 930
\bibitem[1996]{Dubet96}Dubner, G., Giacani, E.B., Goss, W.M., Moffett, D.A., Holdaway, M. 1996, AJ, 111, 1304
\bibitem[2005]{Eiset05}Eisenhauer, F., Genzel, R., Alexander, T. et al. 2005, ApJ, 628, 246
\bibitem[1998]{Fraet98}Frail, D.A., \& Mitchell, G. F. 1998, ApJ, 508, 690
\bibitem[1994]{Fraet94}Frail, D.A., Goss, W.M., Slysh, V.I. 1994, ApJ, 424, 111
\bibitem[1992]{Kaset92}Kassim, N.E. 1992, AJ, 103, 943
\bibitem[2007]{Kotet07}Kothes, R., Dougherty, S.M. 2007, A\&A, 468, 993
\bibitem[2006]{Jaket06}Jackson, J.M., Rathborne, J.M., Shah, R.H. et al. 2006, ApJ Suppl., 163, 145
\bibitem[1990]{Lanet90}Langer, W.D., Penzias, A.A. 1990, ApJ, 357, 477
\bibitem[2006]{Saset06}Sasaki, M., Kothes, R., Plucinsky, P.P., Gaetz, T.J., \& Brunt, C. M. 2006, ApJ, 642, L149
\bibitem[2006]{Stiet06}Stil, J.M., Taylor, A.R., Dickey, J.M. et al. 2006, AJ, 132, 1158
\bibitem[2007]{Tiaet07}Tian, W.W., Li, Z., Leahy, D.A., Wang, Q.D. 2007, ApJ, 657, L25
\bibitem[2003]{Toret03}Torres, D.F., Romero, G.E., Dame, T.M. et al. 2003, Phy. Reports, 382, 303
\bibitem[2002]{Waret02}Wardle, M., Yusef-Zadeh, F. 2002, Science, 28, 2350.
\bibitem[1999]{Waret99}Wardle, M., 1999, ApJ, 525, 101L
\end{thebibliography}
\end{document}